\documentclass[doublecol]{epl2}
\usepackage[utf8]{inputenc}
\DeclareUnicodeCharacter{2009}{\, }
\usepackage[T1]{fontenc}
\usepackage[english]{babel}
\usepackage{amsmath}
\usepackage{amsfonts}
\usepackage{amssymb}
\usepackage{graphicx}
\usepackage[colorlinks=true, citecolor=blue, urlcolor=red]{hyperref}
\usepackage{siunitx}
\usepackage[normalem]{ulem}
\usepackage{xcolor}
\usepackage[numbers,compress]{natbib}
\graphicspath{{./figures/figures_pdf/}}

\newcommand{\vfb}{V_{FB}}
\newcommand{\vdc}{V_{DC}}
\newcommand{\tr}{t_r}
\DeclareMathOperator{\erf}{erf}

\newcommand{\myfig}[2]{
	\begin{figure}[ht]
	\centering
	\includegraphics[scale=0.8]{#1}
	\caption{#2}
	\end{figure}
}

\title{Inertial effects in discrete sampling information engines}

\author{{Aubin Archambault, Caroline Crauste-Thibierge, Sergio Ciliberto, Ludovic Bellon}}

\shortauthor{A. Archambault \etal}

\institute{Univ Lyon, ENS de Lyon, CNRS, Laboratoire de Physique, F-69342 Lyon, France}

\abstract{
We describe an experiment on an underdamped mechanical oscillator used as an information engine. The system is equivalent to an inertial Brownian particle confined in a harmonic potential whose center is controlled by a feedback protocol which measures the particle position at a specific sampling frequency $1/\tau$. Several feedback protocols are applied and the power generated by the engine is measured as a function of the oscillator parameters and the sampling frequency. The optimal parameters are then determined. The results are compared to the theoretical predictions and numerical simulations on overdamped systems. We highlight the specific effects of inertia, which can be used to increase the amount of power extracted by the engine. In the regime of large $\tau$, we show that the produced work has a tight bound determined by information theories.
}

\begin{document}

\maketitle
\nopagebreak
\section{Introduction}
The thermodynamics of feedback controlled systems is a widely studied subject not only for its large number of applications but also for its fundamental aspects~\cite{bechhoefer_control_2021}, which are particularly interesting in mesoscopic systems where thermal fluctuations cannot be neglected. One of these aspects is the connection between information and thermodynamics that goes back to the famous Maxwell demon~\cite{leff_maxwells_1990, lutz_information_2015}. Indeed a mesoscopic machine may produce work from its thermal fluctuations using the information on the system state gathered by the feedback. It has been proved that the produced power $\dot w$ is always bounded by some information acquisition rate $\dot I$, i.e. 
\begin{equation}
	\dot w \le k_B T \dot I,
	\label{eq:W_I}
\end{equation} 
where $k_B$ is Boltzmann's constant and $T$ the bath temperature~\cite{sagawa_generalized_2010, horowitz_nonequilibrium_2010, abreu_thermodynamics_2012, lahiri_fluctuation_2012, rosinberg_continuous_2016, ashida_general_2014}. The first inequality of this form was derived by Sagawa and Ueda for a single feedback loop~\cite{sagawa_generalized_2010}, but subsequently has been extended to include the repeated use of feedback~\cite{horowitz_nonequilibrium_2010, rosinberg_continuous_2016}, allowing for the application to continuously operating information engines. Often $I$ is the mutual information $I_{x, y}$ between the controlled variable $x$ and the outcome $y$ of the measurement performed by the feedback on the system. However this definition doesn't take into account the relevance of the feedback to the measurement. If the protocol is poorly chosen, the measured information $I$ can be high, while the work extraction will be low. {Considering backward processes, Ashida and coworkers~\cite{ashida_general_2014} proposed a modified version of the Sagawa-Ueda equality~\cite{sagawa_generalized_2010}, itself an extension of the Jarzynski equality\cite{Jarzynski-1997}}. They define an unavailable information $I_u$ that is the part of the measured information that is not used to extract work by the chosen protocol. Their generalization of the Jarzynski equality is the following:
\begin{equation}
	\langle\exp{(-{w \over k_BT}-I+I_u)}\rangle =\exp\left({-\Delta F \over k_BT}\right),
	\label{eq:Jar_u}
\end{equation} 
where $\Delta F$ is the free energy difference between the final and initial states. {The definition of $I$ used here is not the mutual information but $I(y)=-\log(P(y)\delta y)$ where $P(y)${$\delta y$} is the probability of getting the measurement outcome within the interval $[y,y+\delta y]$. Instead $I_u=-\log(P_B(y)\delta y)$ relies on the probability $P_B(y)${$\delta y$} of getting in the reverse process the outcome  of the measurements within $[y,y+\delta y]$ and it is accessible experimentally only if the backward process can be performed. The divergence  of $I$ for $\delta y \rightarrow 0$ disappears as  the quantity  $I-I_u$ is  used in Eq.\ref{eq:Jar_u} \cite{ashida_general_2014}.   }

The properties of Eqs.~\ref{eq:W_I}, \ref{eq:Jar_u} have been investigated both theoretically~\cite{horowitz_thermodynamic_2011, abreu_thermodynamics_2012, lahiri_fluctuation_2012, rosinberg_continuous_2016, ashida_general_2014} and experimentally in several different systems~\cite{toyabe_experimental_2010, koski_experimental_2014, koski_-chip_2015, camati_experimental_2016, chida_power_2017, admon_experimental_2018, paneru_lossless_2018, ribezzi-crivellari_large_2019, taghvaei_relation_2022, saha_bayesian_2022}. From the experimental side the results concerned overdamped systems where the role of inertia can be neglected. In this letter we investigate experimentally several protocols that allow us to produce work from thermal fluctuations in an underdamped system where inertia plays a role. The results are compared with theoretical and numerical results in overdamped systems and although we find several analogies, there are features that pertain only to underdamped systems.
\begin{figure}[ht]
	\centering
	\includegraphics[width=0.8\columnwidth]{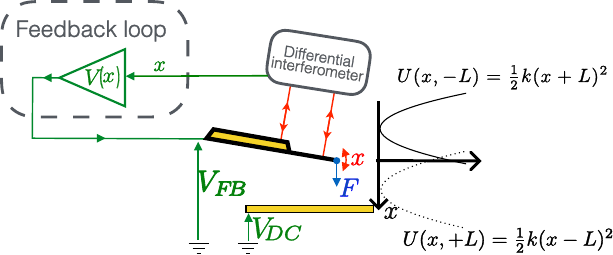}
	\caption{\label{fig:setup}Experimental setup. The deflection of the cantilever $x$ is measured using an interferometer. The deflection is used by the feedback loop to compute a voltage $V_{FB}(x)$ that generates a force on the cantilever, shifting the center of the harmonic potential.}
\end{figure}

\section{Setup} 
In our experiment a conductive micro-cantilever (Mikromotive Octensis 1000S) acts as an underdamped mechanical oscillator submitted to thermal fluctuations. Fig.~\ref{fig:setup} sketches our setup, which is similar to the one described in refs.~\cite{dago_information_2021, dago_virtual_2022}. Specifically the first oscillation mode of the cantilever is used as a underdamped harmonic oscillator characterized by a stiffness $k\simeq \SI{5e-3}{\newton\per\meter}$, a resonance frequency $f_0 = \SI{1087}{\hertz}$, {and a relaxation time $t_r$}. The quality factor {$Q=\pi t_r f_0$} of this oscillator can be tuned by removing the air in the cantilever chamber, from $Q\sim10$ at atmospheric pressure to $Q\sim100$ in light vacuum (\SI{1}{mbar}). The tip deflection $x$ follows the dynamics of a 1D underdamped Brownian particle. The standard deviation of $x$ in thermal equilibrium is $\sigma =\sqrt{k_BT/k}\simeq \SI{0.8}{\nano\meter}$. In the following, we express all lengths in units of $\sigma$ and all energies in units of $k_BT$.

The cantilever deflection is measured by an interferometer~\cite{paolino_quadrature_2013} whose outputs are digitized and sent to a field programmable gate array device (National Instrument FPGA 7975R) that computes the deflection $x$~\cite{dago_virtual_2023}. The device can be programmed to output a feedback voltage $\vfb$ computed using $x$ and a set of rules implemented by the user. It is linked to a computer that records all the relevant data from the experiment.

The feedback voltage, $\vfb$, output by the FPGA is applied to the cantilever. A DC voltage, $\vdc \simeq \SI{90}{\volt}$ kept constant through all experiments, is applied to a conductive flat surface about $\SI{500}{\micro\meter}$ away. This results in a feedback force on the cantilever $F_{FB} $~\cite{butt_force_2005}:
\begin{equation}
	F_{FB} \propto (\vfb - \vdc)^2 = ( \vdc^2 - 2\vdc \vfb + \vfb ^2).
\end{equation}
{The $\vdc ^2$  term is constant throughout an experiment. It shifts the mean equilibrium position of the cantilever when $\vfb =0$. By setting the origin	of the positions, $x = 0$, to the center of the shifted harmonic potential, we can absorb this contribution to the force and remove the term $\vdc ^2$ from the expression of $F_{FB}$.}
Since the maximum voltage $\vfb$ possible for the FPGA is $\SI{1}{\volt}$, $\vfb ^2 \ll \vdc \vfb $ and we can {neglect it in} $F_{FB}$. The resulting expression for the force is $F_{FB} \propto 2\vdc \vfb $. The DC bias acts as an amplification factor used experimentally to tune the sensitivity of the cantilever to the feedback force. 

The delay of the feedback loop is about $1\si{\micro\second}$ which is three order of magnitude smaller than the period of the oscillator $t_0 =\SI{0.92}{ms}$. Thus the feedback is much faster than the oscillator dynamics~\cite{dago_virtual_2023}. Furthermore the use of the FPGA device allows us to run different experiments with different types of feedback without any modification of the experimental setup configuration. 
{Summarizing, the equation of motion  of the cantilever tip submitted to $F_{FB}$ is {a Langevin like equation}:
\begin{equation}
	\ddot{x}+Q^{-1}\omega_0\dot{x}+\omega_0^2x=\omega_0^2\frac{F_{FB}}{k \sigma}+\sqrt{2Q^{-1}} \omega_0 \xi
\end{equation}
where $\omega_0=2\pi f_0$, $x$ is measured in units of $\sigma$ and  $\xi$ is a delta correlated white noise of unit variance.}
\section{Discrete sampling protocol}
\begin{figure}[ht]
	\centering
	\includegraphics[width=0.8\columnwidth]{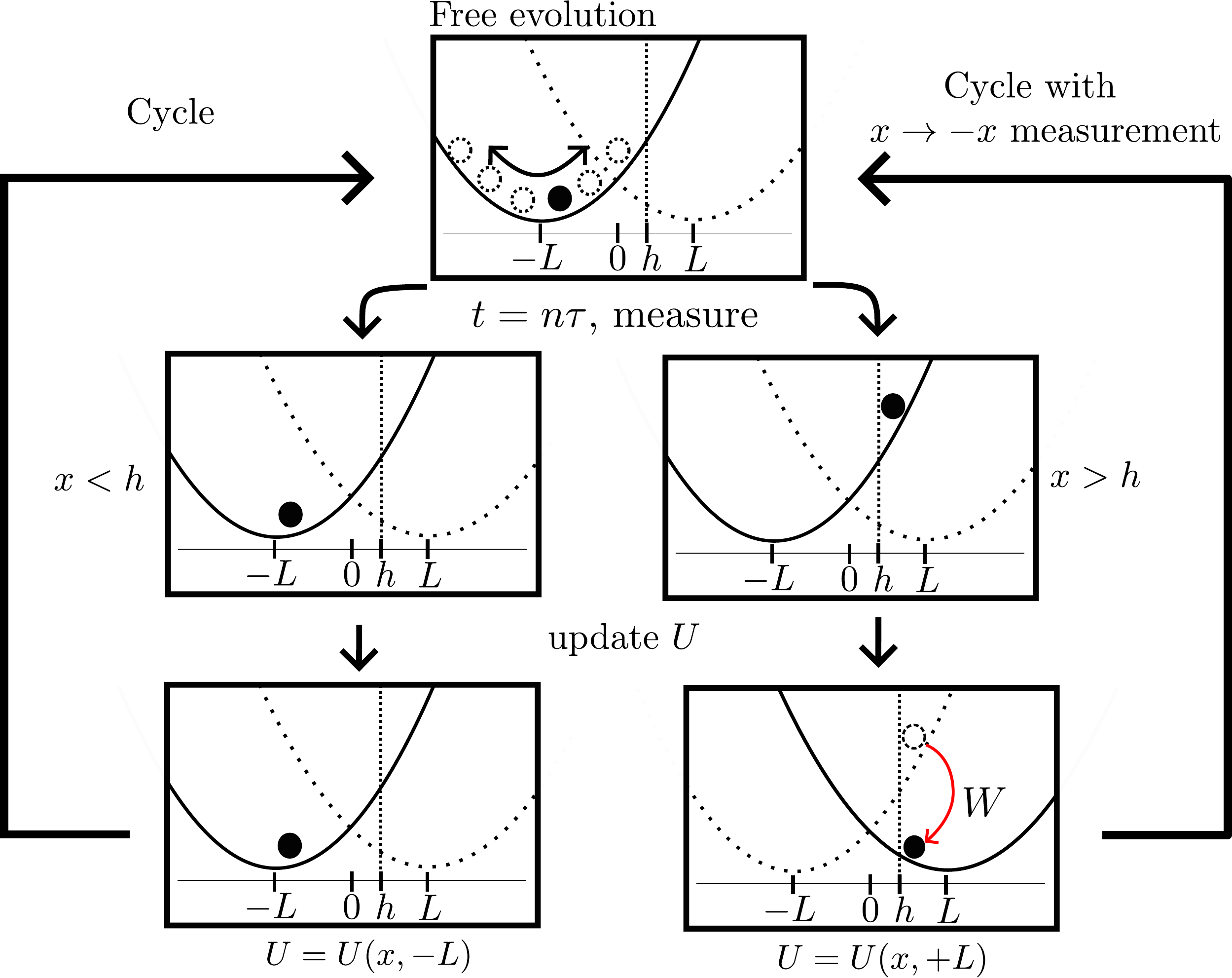}
	\caption{\label{fig:discrete_protocol}The discrete sampling protocol. Starting with a potential $U(x, -L) = \frac{1}{2}(x+L)^2$ centered in $-L$, every time step $\tau$ the position $x$ of the particle is measured. If $x>h$ the potential center is switched from $-L$ to $L$. In both cases we wait a time $\tau$ before performing the next measure.}
\end{figure}
Here we focus on a protocol, illustrated in Fig.~\ref{fig:discrete_protocol}, where the position of the particle $x$ is read by the feedback with a sampling rate $\tau$. Starting from equilibrium with a potential $U(x, -L) = \frac{1}{2}(x+L)^2$ centered in {$-L=F_{FB}/(k\sigma)$}, at each reading the measured $x$ is compared with a threshold $h$. If $x < h$, nothing is done and we wait for a time $\tau$ before performing a new measurement. If $x > h$, the potential is instantaneously switched to $U(x, +L) = \frac{1}{2}(x-L)^2$. During this process, the potential energy of the particle is lowered and the internal energy of the particle decreases. Since the potential is switched almost instantaneously, the heat $q$ exchanged by the particle with the bath and the kinetic energy change $\Delta K$ are both zero at the switching time. Instead the change in the internal energy of the particle is 
\begin{equation}\label{eq:work}
	\Delta U = \frac{1}{2}(x-L)^2 -\frac{1}{2}(x+L)^2 = -2Lx.
\end{equation}
Since $q=0$ and $\Delta K =0$, $\Delta U = -w <0$ if $x>0$. By convention, $w$ is defined as the work performed by the particle on the feedback, thus $w>0$ means that the feedback is extracting work from the particle. The particle is left in the potential centered in $L$ for a time $\tau$. Afterwards a new cycle can be performed symmetrically, switching the potential from $U(x, L)$ to $U(x, -L)$ when $x<-h$. 

This protocol is similar to the protocol proposed by T.Sagawa and M.Ueda~\cite{sagawa_generalized_2010}. The difference between the two protocols is that in our protocol the particle switches back and forth between two potential wells, centered in $\pm L$, whereas the protocol from Sagawa and Ueda uses successive potential wells centered in $-L$, $L$, $3L$, etc. 

\begin{figure}[ht]
	\centering
	\includegraphics[scale=1]{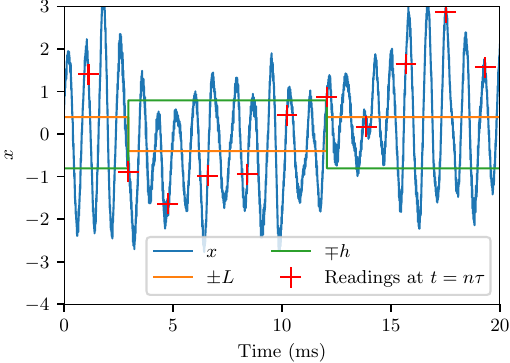}
	\caption{\label{fig:exp_signals} Experimental signals time traces. The blue curve is the measured deflection of the cantilever, $x$. The orange curve is the position of the center of the potential well and takes values $\pm L$. The green curve takes values $\mp h$ and is the threshold the deflection has to cross for the potential to be switched. Red crosses mark instants when the feedback reads the position of the particle. Since the readings happen only every time $\tau$, the particle can cross the threshold without feedback on the potential if it happens between the readings. Experiments with $\tau\simeq 0.8\, \tr=2\, t_0=\SI{1.8}{ms}$, $L=0.4$, $h=0.8$.}
\end{figure}

Experimentally, the position of the particle is measured continuously with a sampling frequency of $\SI{2}{MHz}$ but the feedback is allowed to act only at discrete times with a period $\tau$. Experimental signals are presented in Fig.~\ref{fig:exp_signals}. From the measured trajectory and applied potential, we can compute the work of the particle on the electric field along the trajectory using Stratonovich convention. We can then obtain the statistics of the work performed by the particle each time the potential is switched.

\section{Long times limit}\label{sec:largetau}
\myfig{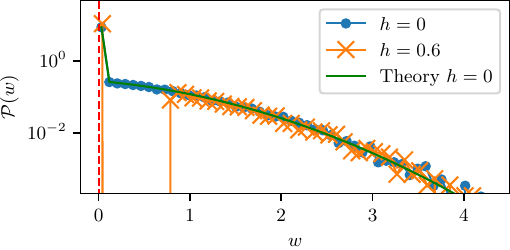}{\label{fig:pdf_no_hyst} Probability distribution functions of the work performed by the particle at each reading of the feedback loop for $L=0.6$ with $h=0$, $h=0.6$ and $\tau \gtrsim 6\, \tr$. The measured pdf is obtained over $10^4$ readings of the position. Events are for $w\geq 0$, which corresponds to work extraction. A strong peak in $w=0$ corresponds to all the readings where $x<h$ and the potential is unchanged, hence $w=0$.}
\begin{figure}[htb]
	\centering
	\includegraphics[scale=1]{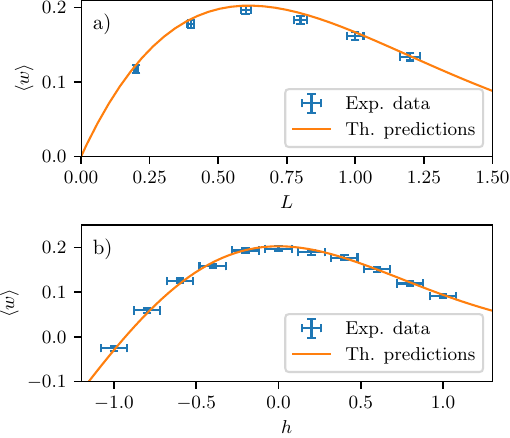}	
	\caption{\label{fig:power_vs_L}a) Mean work $\langle w \rangle$ as a function of $L$ with $h=0$ in the limit $\tau \rightarrow \infty $. An optimum appears for $L=0.6$.
		b)\label{fig:hyst_large_tau} Mean work $\langle w \rangle$ as a function of the position of the threshold $h$ in the limit $\tau \to \infty$, for a fixed value of $L=0.6$. {The error bars represent the statistical uncertainty (one std).} {$L$ and $h$ are in $\sigma$ units.}}
\end{figure}

We first study the limit $\tau \to \infty$. This situation corresponds to the case where the particle is back to equilibrium between each reading of the feedback loop. Experimentally, we take $\tau\gtrsim 6\, \tr$.

At each reading of $x$ by the feedback loop, we compute the work $w$ extracted by the protocol. The probability distribution function (pdf) of $w$ is shown in Fig.~\ref{fig:pdf_no_hyst} for the case $L=0.6$ and $h=0$. As expected, all events correspond to $w\geq0$ and work is extracted by the feedback. A strong peak is present at $w=0$ and corresponds to events where the feedback reads a position $x<h$ and does not switch the potential, leading to zero work. 

The parameters $L$ and $h$ can be chosen to maximize the amount of work and the mean work $\langle w \rangle$ extracted by the feedback at each reading. Fig.~\ref{fig:power_vs_L} shows the extracted work as a function of $L$ at $h=0$: it presents a maximum in $L=0.6$. 
One can also change the position of the threshold $h$, however the optimal value is $h=0$. Indeed, if $h<0$, then we are allowing the system to switch when the particle is in $h<x<0$ with $w<0$ according to Eq.~\ref{eq:work}. In this case, the feedback is providing work to the system to put it in a higher energy state when the potential is switched. If $h>0$, we are imposing a minimal amount of work $w_0=2Lh$ to be extracted each time $x>h$. It means that each time the particle is measured at $0<x<h$, no work is extracted: $w=0$. Conversely, if the potential was switched we would have $w>0$. The feedback is thus loosing occasions to extract work. This has been verified experimentally by measuring the extracted power at different values of $h$ and fixed value of $L=0.6$. The resulting pdf are shown in Fig.~\ref{fig:pdf_no_hyst}. The extracted work as a function of $h$ is shown in Fig.~\ref{fig:hyst_large_tau}. As expected the extracted work decreases as soon as $h\neq 0$. 

Furthermore, we can easily compare the extracted power to theoretical predictions. Indeed, the distribution of the outcome of the reading is described by the equilibrium distribution in an harmonic potential centered in $-L$, $P_{eq}(x, -L) = \exp(-\frac{1}{2}(x+L)^2)/\sqrt{2\pi} $. From this distribution and using eq. \ref{eq:work}, we can deduce the probability distribution of the extracted work: 
\begin{equation}\label{eq:Peq_w}
	\mathcal{P}(w) = 
	\begin{cases} 
		\frac{1}{2}\left(1+\erf\left( \frac{L+h}{\sqrt{2}}\right)\right)\delta(w) &\text{if } w=0, \\
		\frac{1}{2L\sqrt{2\pi}}\exp\left(-{\frac{1}{2}(\frac{w}{2L}+L)^2}\right) &\text{if }w>w_0, \\
		0 &\text{if }w<w_0.
	\end{cases}
\end{equation}
{with $w_0=2Lh$.} This distribution is plotted in Fig.~\ref{fig:pdf_no_hyst} for the case $L=0.6$ and $h=0$ and shows a perfect agreement to our experimental data.
We can also compute the mean extracted work as a function of $L$ and $h$: 
\begin{align}
	\langle w \rangle&= \int_{-\infty}^h 0 \!\times\! P_{eq}(x;-L) dx + \int^{+\infty}_h 2Lx \!\times\! P_{eq}(x;-L) dx \nonumber \\
	&=L^2 \left(\erf\left(\frac{L+h}{\sqrt{2}}\right)-1\right)+L\sqrt{\frac{2}{\pi}}e^{-\frac{1}{2}(L+h)^2},
\end{align}
which gives the theoretical curves in Fig.~\ref{fig:power_vs_L}.

Since the system has time to relax to equilibrium between each reading by the feedback, there is no memory of the previous reading or of the previous switch in the potential. Therefore it is equivalent to operate back and forth between two potential wells or with successive wells. {Furthermore, as all measurements are performed at equilibrium, inertia does not play any role in the limit $\tau\rightarrow\infty$}. In this regime our protocol is equivalent to the one of Ref.~\cite{sagawa_generalized_2010}, for which analytical results have been derived by {Park} and coworkers~\cite{park_optimal_2016}. They compute analytically the average work extracted per event for different values of $L$ and $h$, in the case where $\tau \to \infty $. They find that the largest work is extracted for $L = 0.6$ and $h=0$, which corresponds to the result that we obtain experimentally. 

\section{Intermediate regime}

\begin{figure}[htb]
	\centering
	\includegraphics[scale=1]{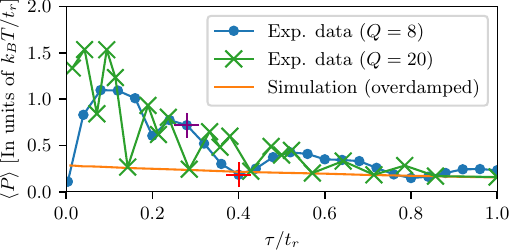}
	\caption{\label{fig:eff_vs_tau}Extracted power as a function of $\tau$ in unit of $t_r$, the relaxation time of the system. In blue, experimental data with $L=0.6$, $h=0$ and $Q=8$ in the underdamped regime. In green, experimental data with $L=0.6$, $h=0$ and $Q=20$. In orange, data from simulations of an overdamped Brownian particle. {The error bars are smaller than the symbols size.} }
\end{figure}

\myfig{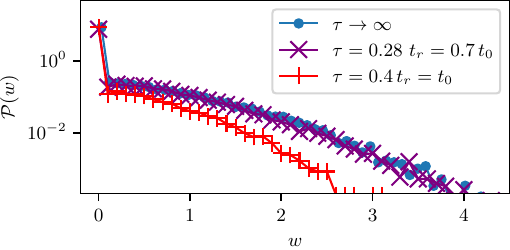}{\label{fig:anomalous_pdf} Probability distribution function of extracted work after readings for $L=0.6$, $h=0$, $Q=8$, with $\tau = 0.28 \, \tr =0.7\, t_0$ (purple) and $\tau = 0.4 \, \tr = 1 \, t_0$ (red), compared with the distribution for $\tau\to\infty$ (blue).}

Operating the engine in a regime of large $\tau$ leads to work extraction, but due to the long time between cycles, the extracted power $P=\langle w \rangle /\tau$ is very low. A way to increase the power output of the engine is to reduce the time between cycles $\tau$ by not letting the system totally relax between cycles. We measure the extracted work while keeping $L = 0.6$ and $h=0$ fixed but varying the sampling time from $\tau = 0.1 t_0$ to $\tau = 3 t_0$. For comparison, simulations of an overdamped 1D Brownian particle under the same feedback protocol are performed. We simulate the trajectories of overdamped particles for a potential switching between $U(x, -L)$ and $U(x, +L)$, and compute the work extracted for different values of $L$ and $h$. The simulations are performed using a standard Euler scheme, simulating an overdamped Langevin equation, {with viscosity $\gamma$, stiffness $k$, and relaxation time $t_r=\gamma/k$.} {The simulation starts with a position sampled from equilibrium distribution.} Its duration is typically $100\, \tr$, the first $5\, \tr$ are discarded {to reach a steady state}, and the time step of the simulation is $dt=0.004\, \tr$. The change of the potential is allowed only every time interval $\tau$ which is changed between 0 and $t_r$. Trajectories from the simulation are then analyzed similarly as the experimental ones. To compare the two results, we use $\tr$ as the unit of time, which is different in the two systems. In the underdamped, inertial, regime $\tr = Q/(\pi f_0)$ {while in the overdamped regime $t_r =\gamma/k$}. Figure \ref{fig:eff_vs_tau} shows the comparison of the power extracted for the two regimes. Using $\tr$ as a unit for times, the curve of the overdamped regime appears {phenomenologically} as a lower enveloppe for the curve of the underdamped regime {within experimental errors}. The green curve in Fig.~\ref{fig:eff_vs_tau} shows results from an experiment with higher quality factor $(Q=20)$, obtained by lowering the pressure in the experiment, confirming that $\tr$ is indeed the correct time scale. The extracted power in the underdamped regime is always higher than for the overdamped. {Indeed  the characteristic time for exploring the potential is $t_0$  in underdamped systems  and $\tau_r$ in overdamped. As $t_0 \ll t_r$  then   $h$ is reached faster in underdamped than in overdamped systems.   }

However, anomalies on the curve appear for the underdamped regime at specific values of $\tau$, where the extracted power drops. To understand these anomalies, we plot on Fig.~\ref{fig:anomalous_pdf} the pdf of the extracted work for different values of $\tau$: $0.4\, \tr$ and $0.28\, \tr$ corresponding respectively to the red and purple crosses on Fig.~\ref{fig:eff_vs_tau}. Two effects can be noticed. First the peak in $w=0$ is stronger {(by 10\%)} for the anomalous case. This means that more measurements have to be performed before the particle is found on the right side of the threshold. The second effect is that in the anomalous case, the spread of the non-zero part of the pdf is smaller. Since $w = 2Lx$, this means that the particle is found closer to the threshold when the measurement is performed. Both effects contributes to a reduction of the extracted power and can be understood as a synchronisation effect between our measurements and the natural oscillation of the underdamped particle. Indeed, the anomalies of Fig.~\ref{fig:eff_vs_tau} fall exactly on integer and half-integer values of $t_0$, the period of the oscillator, specifically in the example of Fig.~\ref{fig:eff_vs_tau} at $Q=8$, $\tau= t_0 \simeq 0.4\, \tr$.

\section{Short times limit}\label{sec:tau0}
\myfig{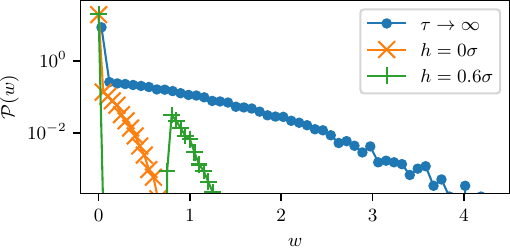}{\label{fig:histograms_tau0}Histograms of work measured at each reading for $\tau=0.004 \, \tr$, $L=0.6$ and $h=0$ (orange curve) and $h=0.6$ (green curve). Since $\tau \ll \tr$, the potential is switched soon after the particle reaches the threshold, which leads to much narrower distributions compared to the case $\tau\to\infty$ from figure \ref{fig:pdf_no_hyst} (reproduced here in blue). The effect of moving the threshold $h$ is that the position of the histogram is switched from around $0$ to a non-zero value. }

\begin{figure}[htb]
	\centering
	\includegraphics[scale=1]{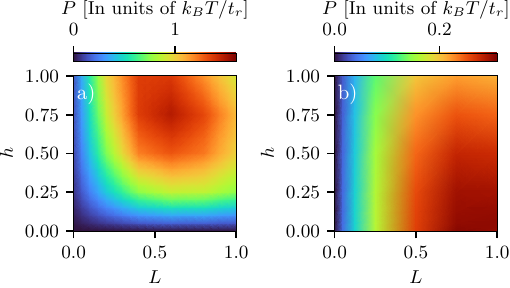}
	\caption{\label{fig:map_exp}a) Experimental heatmap of the extracted power as a function of $L$ and $h$ for $\tau = 0.004\, \tr$ in the underdamped regime. The optimal value is at $L=0.6$ and $h=0.75$.
		\label{fig:map_simu} b)~Computed heatmap of the extracted power as a function of $L$ and $h$ for $\tau=0.004\, \tr$ in the simulated overdamped regime. The optimal value is at $L=0.75$ and $h=0$. }
\end{figure}

On Fig.~\ref{fig:eff_vs_tau}, we can notice that the power goes to $0$ as $\tau\to 0$ {for the dataset $Q=8$. We expect the same behavior for $Q=20$, but with a sharper drop and we only see the beginning of the decay with the shortest $\tau$ we used.} To understand this phenomenon, we did experiments in the regime where $\tau \ll \tr $.
This regime, in which $x$ is sampled continuously, can be studied as a first-passage problem of a particle in the potential $U(x, -L)$ [or in $U(x, L)$], whose trajectories start at $-h$ [$h$], as a result of the previous potential switch, and end at $h$ [$-h$]. Since the comparator reads the position of the system continuously, the excursion of the particle above the threshold is limited and the potential is switched as soon as the particle reaches $x=h$. The work extracted is then $w=2Lh$ and we can immediately see that if $h=0$, no work can be extracted as shown by the pdf of the work in Fig.~\ref{fig:histograms_tau0}. If the threshold is moved to $h>0$, the work extracted will be non-zero. However, as $h$ is raised, the position of the target threshold is moved away from the initial position, and the time needed for the particle to reach the threshold increases. As a consequence there is an optimal value of $h$ that has to be found.

An experimental map of the power extracted as a function of $L$ and $h$ is displayed in Fig.~\ref{fig:map_exp}. An optimum for the extracted power is found at $L=0.6$ and $h=0.75${, giving an extracted power $P=1.48\,k_BT/\tr$}. To compare with the overdamped case, we use the same simulation technique as for the intermediate time regime, taking $\tau = dt = 0.004\, \tr$ for consistency with the experiments. From these simulations, we can compute a map of the extracted power as we do for the experiments. The resulting heatmap is shown in Fig.~\ref{fig:map_simu}. 

A huge difference can be seen between the two regimes. In the overdamped case, the heatmap is dominated by an increase in the extracted power at large $L$ with light dependence on $h$. Optimal work extraction is achieved for $L=0.75$ and $h=0$, giving an extracted power of $P=0.3\,k_BT/\tr${, smaller than the optimum in the underdamped regime}. The main difference between the two results is that while in the underdamped regime it is very efficient to work with $h\neq0$, the overdamped regime is optimal when $h=0$. This can be understood as in the underdamped regime the dynamic is ballistic, thus the particle crosses $0$ with some velocity that will allow the particle to reach a non zero threshold. In the overdamped regime, the viscosity prevents such guesses on the future position of the particle and the optimal is for a threshold in $h=0${ from the same argument as for $\tau\to\infty$}. 

As a conclusion, at small $\tau$ the power can be optimized by tuning the values of $L$ and $h$ and in the underdamped case this power becomes very large. It is also interesting to notice that in the overdamped case the largest power is obtained at $L=0.75$ and $h\to 0$ which is very different from the values of~\cite{park_optimal_2016}. It is important to point out that this difference comes from the fact that our system oscillates from $-L$ to $L$ whereas their system translates always in the same direction.

\section{Information}
\begin{figure}[htb]
	\centering
	\includegraphics[scale=1]{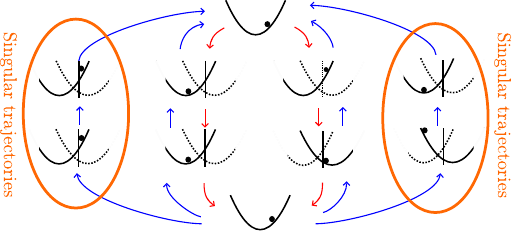}
	\caption{\label{fig:sing_trajs} The four kinds of trajectories in our experiments. Two of them are possible in the forward process (red arrow), and two are singular and only possible in the backward process (blue arrow). These singular trajectories correspond to cases where the particle is above the threshold and the potential is unchanged, or where the particle is below the threshold and the potential is changed. }
\end{figure}
\myfig{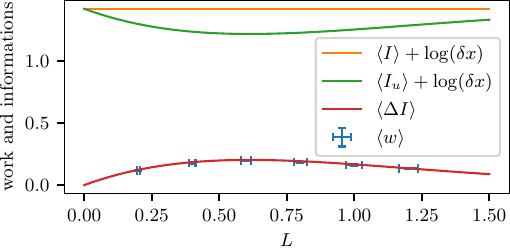}{\label{fig:info} Blue: $\langle w \rangle$ measured experimentally as a function of $L$ in the regime of large $\tau$, with $h=0$. Orange: Theoretically predicted information {$<I> +\log (\delta x)$}. Green: Theoretical prediction for unavailable information {$<I_u>+\log (\delta x)$.} Red: Comparison with the predicted $<\Delta I >\equiv <I -I_u >$. }

{For the case of $\tau \to \infty$, information can be measured using the method in ref.~\cite{ashida_general_2014}, where}
{ the notion of unavailable information is introduced to give a tighter bound on the work extracted  from information engines.  This quantity  estimates  the amount of information that is measured from the system but unused by the protocol.}

This unused information can be computed by studying the time reversal of the process. In the forward protocol, the outcome $m$ of a measurement at time $t_m$ imposes the protocol $\lambda$, which depends on the measurements only at time $t>t_m$. However, in the reverse process, the protocol depends on $m$ at times $t<-t_m$, before the measurement. Due to causality a feedback cannot be performed and the protocol has to be imposed beforehand. This makes some trajectories, that are forbidden in the direct process, possible in the reverse. Such trajectories are called singular trajectories.

For our protocol, there are four kinds of trajectories presented in Fig.~\ref{fig:sing_trajs}. 
The proposed way of computing the unused information $I_u$ in~\cite{ashida_general_2014} is based on computing the probability of the non-singular trajectories for a given measurement outcome.

 {More specifically our feedback measures  the position of the particle $x$ with a negligible error $\delta x$, thus the extracted information in the forward process is: $I(x)=-\log(P_{eq}(x;-L) \delta x)$.  The orange curve in Fig.~\ref{fig:info} shows  that $\langle I\rangle +\log(\delta x)$ is independent of the protocol used and is therefore constant in  our experiments for all $L$ and $h$.}

{In the backward process  one has to consider the potential in which the backward trajectories start. If the outcome of the forward measurement is $x<h$, then the potential for the backward trajectory is unchanged and $I_u(x) = -\log(P_{eq}(x;-L)\delta x)$. Instead if the outcome of the forward measurement is $x>h$, then the potential for the backward trajectory is changed and $I_u(x) = -\log(P_{eq}(x;+L)\delta x)$. We notice that  $\langle I_u \rangle+\log(\delta x)$ depends on $L$, as illustrated by the green curve on Fig.~\ref{fig:info}.} 
{We also  notice that for $x<h$, $w=0$ and $\Delta I = I-I_u=0$; whereas for $x>h$, $w=2Lx$ and $\Delta I=2Lx$. As a consequence $w-\Delta I =0$ for all measurement outcomes and the {extended Sagawa-Ueda} equality, $\langle e^{w-\Delta I}\rangle = 1$, is satisfied. } 

 Fig.~\ref{fig:info} shows in blue the experimentally measured $\langle w \rangle$ as a function of $L$ for $\tau = 6\, \tr $ and $h=0$ (same data as Fig.~\ref{fig:power_vs_L}). We plot in red  $\langle\Delta I\rangle$ as a function of $L$. We get that $\langle\Delta I\rangle$ is a tight bound for $\langle w \rangle$ thus the inequality $\langle w \rangle \leq \Delta I$ is saturated.  While our protocol  doesn't use all the measured information, no energy is lost by our realization of the protocol and we achieve the maximal work extraction for this given protocol.

In this sense we achieve a lossless information engine~\cite{paneru_lossless_2018}, an engine in which all the used information is converted into extracted work. {It may be possible to imagine a protocol that extracts the same information $I$ from the measurement but has a smaller $I_u$,} {hence being more selective on the possible trajectories allowed. Such protocol should extract more work, and could potentially be optimized for this goal.}

\section{Conclusion}

We have studied experimentally the behavior of an underdamped harmonic oscillator in presence of thermal noise and a feedback that acts as a demon. Thanks to the demon the harmonic oscillator behaves as an information engine able to produce work from the information gathered by the demon on the state of the system. Several feedback protocols have been applied by tuning the system parameters and the feedback sampling time. 
We find the optimal choice of the parameters for the underdamped and overdamped oscillators. We show that the relaxation time is the good timescale to compare the two regimes and that the overdamped regime acts as a lower bound for the underdamped regime. Specific synchronisation effects arise when the sampling time is a multiple of the frequency of the underdamped oscillator, making the choice of the sampling time a key parameter when operating underdamped information engine.
In the limit of very short sampling times, we highlight strong differences in the optimal parameters for the two regimes.
Finally we find that in the limit of very large sampling time the produced work is bounded by the difference between the measured information, which depends on the measurement, and the unused information, which depends on the feedback performed. The same bound is much more difficult to obtain for finite $\tau$ and will be the subject of future investigation. While we achieve maximal work extraction for our chosen feedback, it might be possible to increase the efficiency of the engine by using the same measurement with a different feedback, thus lowering the amount of unused information.

\vspace{0.5 cm}
\textbf{Acknowledgments} 
We thank Salambô Dago for the initial programming of the acquisition setup, as well as Rafna Rafeek, Alberto Imparato and Christopher Jarzynski for insightful scientific discussions. This work has been supported by project ANR-22-CE42-0022. The data supporting this study are available here \cite{Archambault_2024_Dataset}.

\bibliographystyle{apsrev4-2-titles}
\bibliography{refs.bib}

\end{document}